\documentclass{raa_twocolumn}

\usepackage{graphicx,times}             
\usepackage{natbib}
\usepackage{amssymb,amsmath}
\bibpunct{(}{)}{;}{a}{}{,}
\usepackage[pagebackref=true]{hyperref}

\begin{document}

  \title{The Microchannel X-ray Telescope on board the SVOM mission: in-flight scientific performance
}

   \volnopage{Vol.0 (202x) No.0, 000--000}      
   \setcounter{page}{1}          

   \author{D. G\"otz 
      \inst{1,*}\footnotetext{$*$Corresponding Authors, these authors contributed equally to this work.}
    \and S. Crepaldi
\inst{2}
   \and E. Doumayrou
\inst{1}
\and C. Feldman
\inst{3}
   \and P. Ferrando
\inst{1}
 \and A. Fort
\inst{2}
\and H. Goto
\inst{4,1}
\and J. Jaubert
\inst{2}
\and J.-M. Le Duigou
\inst{2}
\and P. Maggi
\inst{5}
\and A. Meuris
\inst{1}
\and M. Moita
\inst{1}
\and K. Mercier
\inst{2}
\and F. Robinet
\inst{6}
\and C. Plasse
\inst{1}
\and A. Sauvageon
\inst{1}
\and C. van Hove
\inst{6}
}

  \institute{AIM-CEA/DRF/Irfu/Departement d’Astrophysique, CNRS, Universite Paris-Saclay, Universite Paris Cite, Orme des Merisiers, Bat. 709, Gif-sur-Yvette, 91191, France; {\it diego.gotz@cea.fr}
  \and
Centre National d’Etudes Spatiales, Centre spatial de Toulouse, 18 avenue Edouard Belin, 31401 Toulouse Cedex 9, France;
\and
Space Park Leicester, University of Leicester, United Kingdom;
\and
College of Science and Engineering, School of Mathematics and Physics, Kanazawa University, Kakuma, 9201192 Kanazawa, Japan;
\and
Observatoire Astronomique de Strasbourg, Université de Strasbourg, CNRS, 11 rue de l’Universite , 67000 Strasbourg, France;
\and
Universit\'e Paris-Saclay, CNRS/IN2P3, IJCLab, 91405 Orsay, France;
}
\vs\no

\abstract{ The Microchannel X-ray Telescope (MXT) is a compact and lightweight focusing X-ray telescope, which is part of the space
payload of the SVOM mission. The main goal of the MXT instrument is to precisely localize and physically characterize the early
phases of the X-ray afterglows detected by the SVOM ECLAIRs coded mask telescope after a satellite slew.
The MXT is composed by a "Lobster-Eye" type optics, with a 58$\times$58 arcmin$^{2}$ field of view,  based on micro-pores of 40 $\mu$m side. This innovative type of optics is coupled to an X-ray camera, which implements at its focal plane a low-noise pnCCD. The MXT system is completed by an onboard calculator, able to command the whole telescope and to analyze in real time the MXT data stream and hence to localize the sources within the MXT field of view.
In this paper, we present the MXT design and in-flight performance, as measured during the SVOM Commissioning and early science operation phase. In particular, we will focus on the optical and spectral performances, the in flight localization capabilities, and how these compare with the pre-flight ground measurements.
\keywords{instrumentation --- X-ray telescopes; gamma-ray bursts; space missions}
}

   \authorrunning{D. Götz et al. }            
   \titlerunning{The MXT Telescope on board the SVOM mission}  

   \maketitle

%
%
\section{Introduction: MXT within the SVOM mission}           
\label{sect:intro}
The Space-based Variable astronomical Object Monitor (\textit{SVOM}; see Cordier et al. this issue) is a Sino-French mission developed, in cooperation by the Chinese National Space Agency (CNSA) and the French Space Agency (CNES). 
The SVOM mission has been designed for the study of Time Domain Astrophysics (TDA) and in particular Gamma-Ray Bursts (GRBs). GRBs are short flashes of gamma rays lasting from less than a second to a few hundreds of seconds, appearing from unpredictable directions over the entire sky. They have been discovered in the late 1960s \citep{klebesadel73}, and have remained a mystery up to the discovery in the 1990s of the so-called \textit{afterglows}, i.e. the electromagnetic emission following the GRBs at other wavelengths (X-ray, optical, IR, radio), lasting a few hours up to several months after the event. 
The first observations of GRB afterglows took place in X-rays, and have been performed for by the Italian
satellite Beppo\textit{SAX} \citep{sax}. These crucial observations allowed the astronomers to precisely localize the GRB counterparts (at $\sim$ arcmin level), and hence permit the identification of their optical counterparts and in fine their host galaxies. This allowed in turn to measure the GRB distances through spectroscopic observations, and finally confirm their cosmological origin  \citep{costa97, vanparadijs97, frail97, metzger97}, putting an end to a three decades long debate about the cosmological nature of GRBs.

The Beppo\textit{SAX} breakthrough made it clear to the scientific community that X-ray afterglow observations using focusing X-ray telescopes play a key role in localizing, studying and understanding the GRB phenomenon.

In fact, since the launch of the Neil Gehrels Observatory (aka \textit{Swift}, \citealt{swift}) in 2004, GRBs are routinely discovered and followed up from ground-based telescopes, thanks to the rapid and precise X-ray localizations provided by it's on board focusing X-Ray Telescope (XRT). Hence it was clear that the SVOM mission needed a focusing X-ray telescope on board to enhance the possibility of making significant progress in GRB studies. 
However, given the limited resources on board \textit{SVOM} wrt. \textit{Swift}, we needed to come up with a design that would provide sufficiently interesting performance, keeping in a limited volume, power and mass allocation. These constraints resulted in the MXT coupling of a narrow-field "Lobster-eye" optics with a state of the art pnCCD, allowing for imaging, timing and spectroscopy in the 0.2--10 keV energy band, with a goal of obtaining sub-arcminute localization less than 10 minutes after the GRB.

The details of the MXT design will be presented in Section \ref{sec:mxt}, while Section \ref{sec:perf} will describe the MXT performance as measured during the SVOM Commissioning and Performance \& Verification phases, which lasted from July 2024 to April 2025. 
More details about the different subsystems can be found in accompanying papers: Moita et al. for the MXT Camera, Feldman et al. for the Optics, and Robinet et al. for the onboard scientific processing. Details about the on ground MXT data analysis pipeline can be found in the paper by Maggi et al.

\section{The MXT Instrument}
\label{sec:mxt}

The SVOM satellite has been launched successfully on June 22$^{nd}$ on a low Earth orbit with an inclination of 29$^{\circ}$ and an altitude of $\sim$600 km.  The SVOM mission is composed of a space segment, as well as a few ground based dedicated follow-up facilities (Wei et al. this issue). The space segment is composed of four co-aligned instruments.
Two instruments (ECLAIRs and GRM) are sensitive in the hard-X/soft gamma-ray energy range and have wide fields of view, in order to monitor vast regions of the sky and detect gamma-ray transients. Two narrow field of view instruments (MXT and VT) are used to follow-up and characterize the afterglow emission.

ECLAIRs (Godet et al. this issue) is a coded-mask telescope, with wide field of view of about 2\,sr (89$^{\circ}\times$89$^{\circ}$), sensitive in the 4\,keV -- 150\,keV energy range, and it comprises an on-board software (Schanne et al. this issue) to detect and localize (to better than 13\,arcmin) in near-real-time the GRBs that appear in its FOV. Once a new transient is detected ECLAIRs issues an alert and requests the platform to slew so that the error box can be observed by the narrow-field instruments. 

ECLAIRs is complemented by the Gamma-Ray Monitor (GRM, Dong et al. this issue), a set of three 1.5 cm thick NaI scintillators of 16 cm in diameter, each one offset by 120$^{\circ}$ wrt. each other and with a combined FOV of about 2.6 sr. The GRM has poor localization capabilities (He et al. this issue), but it has the important merit of extending the SVOM spectral range up to about 5\,MeV, and it also increases the probability of simultaneous detection of short GRBs and gravitational waves alerts, issued by ground based laser inteferometers.

The Visible Telescope (VT, Qiu et al. this issue) is a Ritchey-Chretien telescope with a 40\,cm diameter primary mirror. Its field of view is 26$\times$26 arcmin$^{2}$ wide, adapted to cover the ECLAIRs error box in most of the cases. It has two channels, a blue one (400--650 nm) and a red one (650--1000 nm), and a sensitivity limit of $M_{V}=22.5$ in 300 s, making possible the detection of about 80\,\% of the ECLAIRs GRBs.
The space segment is completed by the MXT, that is described in detail hereafter.

The Microchannel X-ray Telescope has been developed under the CNES responsibility in close collaboration with CEA-Saclay/Irfu , the Univeristy of Leicester, the Max Planck Institut f\"ur Extraterrestrische Physik (MPE) in Munich, and the IJCLab in Orsay. It is a light ($<$\,42\,kg) and compact (focal length $\sim$\,1.12\,m) focusing X-ray telescope; its sensitivity below 1\,mCrab makes it the ideal instrument to detect, identify and localize down to the sub-arc min level X-ray afterglows of the SVOM GRBs.

\begin{figure*}[ht]
   \centering
   \includegraphics[width=0.8\textwidth]{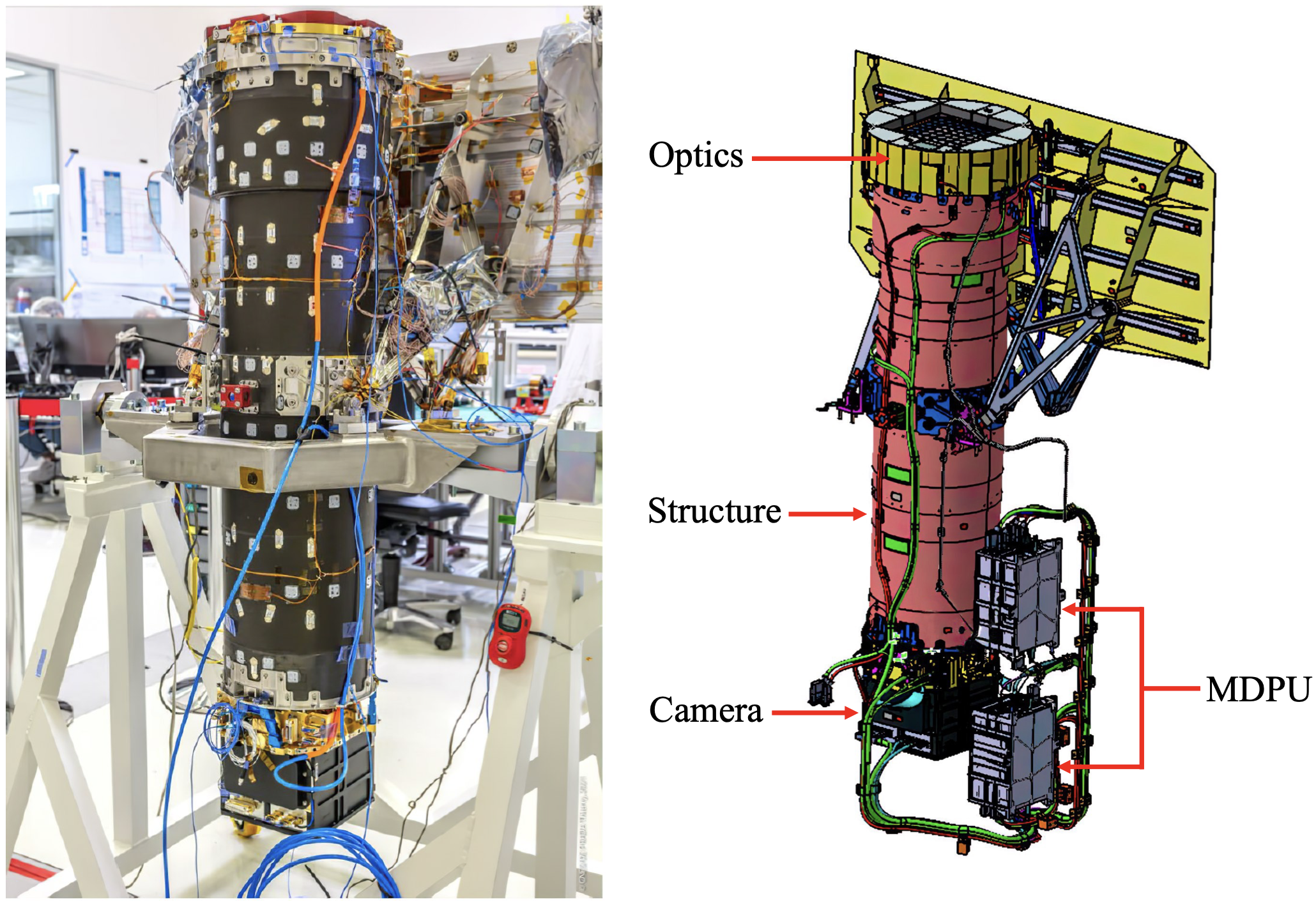}
   \caption{Main components of the Micro-channel X-ray Telescope. (Left) Picture of the integrated (without multi-layer insulator) MXT flight model instrument. (Right) Schematic view of the instrument, showing the optics, the telescope structure, the camera and the two calculating units (MDPUs) (Credit: CNES).}
   \label{fig:mxt}
   \end{figure*}

The MXT (see Fig. \ref{fig:mxt}) is composed of five main sub-systems:
\begin{itemize}
    \item the MOP: the MXT OPtical assembly, based on square Micropore Optics (MPO),
    \item the MCAM: the MXT CAMera hosting a pnCCD,
    \item the MST: the MXT carbon fiber STructure,
    \item the MDPU: the MXT Data Processing Units (in cold redundancy),
    \item the MXT radiator to dissipate the heat generated at focal plane level.
\end{itemize}

\subsection{Optics}

\begin{figure}[ht]
   \centering
\includegraphics[height=4cm]{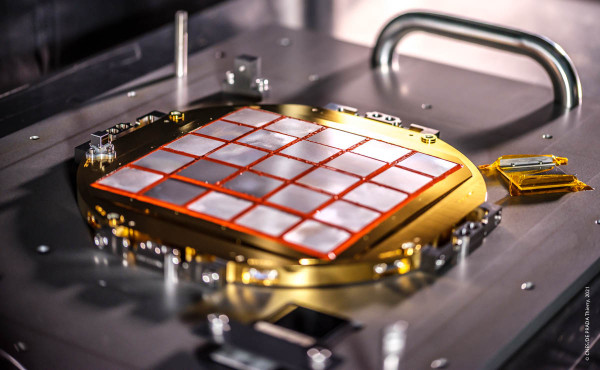}
\includegraphics[height=4cm]{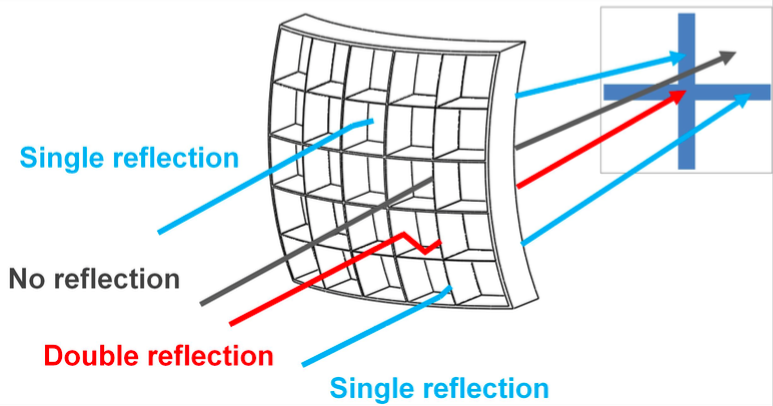}   \caption{(Top) Picture of the MXT optics module (MOP) (source: SVOM website). (Bottom) Schematic of the redirection of X-rays from a distant source by the Lobster eye optics onto a focal plane with a characteristic cruciform pointing spread function (Source: Einstein Probe website).}
   \label{fig:fmmop}
   \end{figure}

The MOP design is based on a ‘‘Lobster-Eye'' grazing incidence X-ray optics, first proposed by \citet{angel79}, and inspired by the vision of some crustacean decapods. It is composed of 25 square MPO plates of 40 mm each arranged in a 5$\times$5 configuration, see Fig.\ref{fig:fmmop}.

Although Lobster-Eye optics have been originally developed for large field of view telescopes (several tens of square degrees), the MXT optical design is optimized for a (relatively) small field of view\footnote {The FOV of the whole array is 6$^{\circ}\times$6$^{\circ}$, and the effective FOV is limited by the detector size to 58$\times$58 arcmin$^{2}$} by making use of a combination of 1.2 and 2.4 mm thick plates (with a pore size of 40 $\mu$m), whose inner walls are coated with Ir to enhance reflectivity. The  entrance of the MPOs pores is covered with a 70 nm thick Al film, and the MPOs are then slumped to match a spherical surface which provides the requested X-ray focusing. This technique results in a peculiar point spread function (PSF), made by a central peak and two cross arms, see Fig. \ref{fig:fmmop}. The central peak is due to photons that are reflected twice on adjacent walls, while the cross arms are due to photons being reflected only once. Finally, a small part of the photons do not interact at all with the optics material and produce a diffuse background.

Despite the relative imaging complexity, the MXT MOP is very well adapted to GRB studies, for which the X-ray afterglow will be during the first minutes of observation the only bright source in the MXT FOV.  Although less performing than classical X-ray Wolter-I type optics, that could be produced to match a 1$^{\circ}$ FOV, the MOP is very light ($<$ 2 kg, more than an order of magnitude lighter wrt. equivalent Wolter-I optics) making lobster-eye optics very attractive for small space borne instruments, like MXT.

\subsection{Camera}

\begin{figure*}
   \centering
\includegraphics[width=0.8\textwidth]{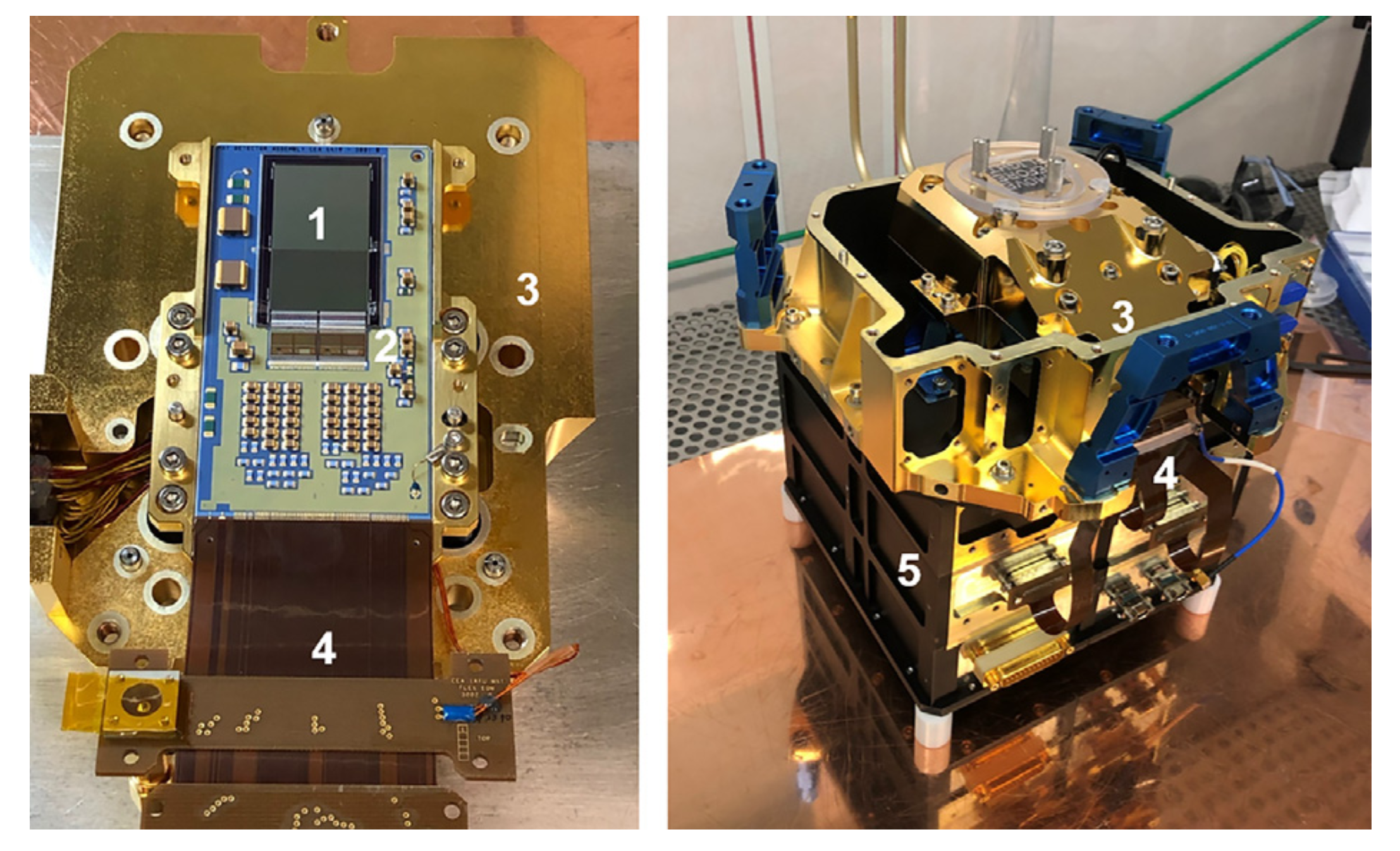}
   
\caption{Focal plane assembly (FPA) of the MXT camera. A ceramic board equipped with a pnCCD (1) and 2 CAMEX ASICs (2) is mounted in a proton shielding (3) and the flex lead (4) is connected to the warm front-end electronics (5) (Source : CEA/DAp).}
   \label{fig:mcam}
   \end{figure*}

Coupled to the MOP is the MXT Camera (MCAM) which implements a focal plane assembly based on a pnCCD \citep{meidinger06} with its readout and control electronics (FEE), see Fig. \ref{fig:mcam}. It also comprises a filter wheel, radiation shielding material and a thermo-electric cooling system \citep{meuris23}.
The detector has an active area of 256$\times$256 square pixels of 75\,$\mu$m side length, and a reduced frame store area with $75\times51~\mu\mathrm{m}^2$ pixels. Once transferred to the frame store area the charges are collected column-wise by two dedicated ASICs called CAMEX. 
This CCD is fully depleted (450\,$\mu$m depth), and its readout rate is 10 frames per second. The detector is actively cooled to $-$65$^{\circ}$C, in order to guarantee a low thermal noise over the nominal 0.2--10 keV energy range, and to reduce the radiation damage effects in flight. The filter wheel allows placing an $^{55}$Fe calibration source or a shutter in front of the detector when needed. Two open positions are present on the filter wheel, but one of them is identified as FILTER. This is due to the fact that the original design
included a supplementary optical filter in order to reduce the optical loading on the detector
due to bright optical sources and Earth albedo. However the flight filters did not survive
the pre-launch shock tests, and it was decided to remove the filter to avoid the risk
of breaking it during launch and hence contamination of the focal plane. Unfortunately this had consequences
on the MXT availability over the mission, since the detector is not able to deal with the increased flux
from the Earth albedo, see \ref{sec:straylight}.

\subsection{MDPU}

The MDPU is responsible for the MXT thermal control, the calibration wheel control, the generation of the telemetry and the handling of the MXT telecommands. Its scientific partition deals with the processing of ‘‘dark frames'' for onboard offset and noise calculations, which are then used to select the pixels transmitted to the ground, i.e. those whose deposited charge exceed a pre-defined threshold (see \citealt{schneider23} for more details). In addition the MDPU scientific partition is responsible for analyzing the MXT data stream on board, by identifying valid X-ray patterns,
building sky images, and detecting and localizing afterglow candidates in near-real-time. The afterglow candidate positions are improved during the observation, as more data are accumulated, and regularly transmitted to ground (every $\sim$\,30 s) in order to allow ground based robotic telescopes to look for optical afterglows in a more efficient way. In fact, the MXT sky error areas will be on average ten times smaller than the ones produced by ECLAIRs, highly enhancing the chances for the VT and other telescopes to correctly identify the optical GRB counterparts. The optical identification and the subsequent spectroscopic distance measurement being a critical task in GRB science, the driving requirement for the MXT design was to be able to localize 90\,\% of the GRBs pointed to after a slew to better than 2 arc min (J2000).

\section{The MXT scientific Performance}
\label{sec:perf}

The MXT has been designed to fulfill the scientific performance presented in Tab. \ref{tab:perfo}.
   \begin{table}[ht]
    \centering
    \caption{MXT expected scientific performance.}
    \renewcommand{\arraystretch}{1.25}
        \begin{tabular}{ll}
            \hline \hline
            Energy range             & $0.2 - 10$~keV \\
            Field of View            & $58 \times 58$~arcmin \\
            Angular resolution       & 10~arcmin at 1.5~keV \\
            Source location accuracy & $< 120$~arcsec for 80\,\% GRBs \\
            Effective area           & $\sim 35$ cm$^2$ at 1.5~keV \\
            Sensitivity ($5 \sigma$) & 10~mCrab in 10~s \\
                                     & 150~$\mu$Crab in 10~ks \\
            Energy resolution        & $<80$ eV at 1.5~keV \\
            Time resolution          & 100~ms \\
            \hline
        \end{tabular}
    \label{tab:perfo}
    \end{table}

Its scientific performance has been evaluated on ground through extensive testing a the MPE Panter facility. The results of these tests can be found in \citet{gotz23}. Here we will focus on the observations early during the SVOM mission that allowed us to confirm the expected MXT performance.

\subsection{The Optical Performance}

\begin{figure}[ht]
   \centering
\includegraphics[height=6cm]{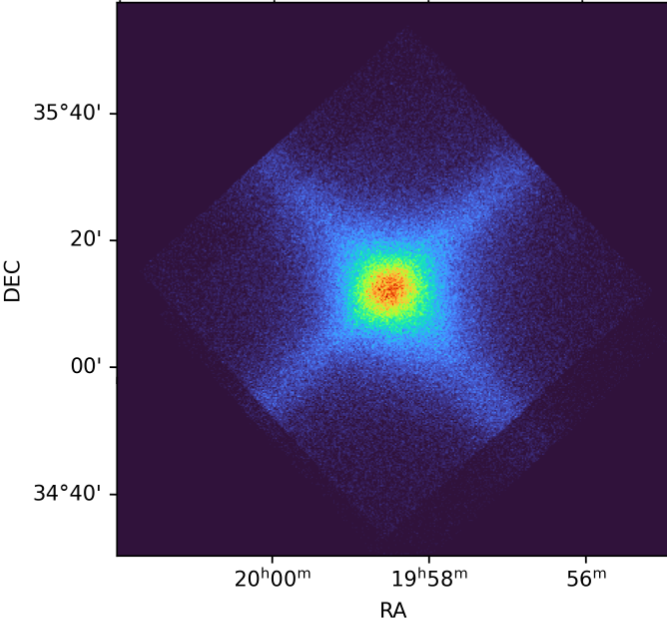}
\includegraphics[height=6cm]{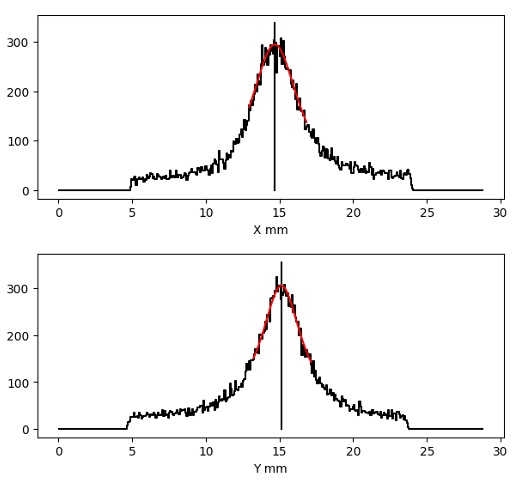}   \caption{(Top) MXT 0.2-10 keV sky image of the field of Cyg X-1. (Bottom) 1D Projection of the MXT image over the two detector axes. The abscissa units are counts, the red line represents a fit obtained with Eq. \ref{eq:opt}.}
   \label{fig:opt}
   \end{figure}

In order to model the MXT point spread function (PSF) we typically use a Lorentzian function of the form

\begin{equation}
\label{eq:opt}
f(x)=\frac{1}{1+(\frac{2x}{G})^2}+c,
\end{equation}

where c represents the contribution of the cross arms\footnote{c depends on the energy and can reach 10\% of the peak value.}, and G the typical width of the PSF (similar to its FWHM). Using Cyg X-1 pointed observations at the beginning of the Commissioning phase, we extracted the photons around 1.5 keV, we obtain G=11.1 arcminutes (see Fig. \ref{fig:opt}), 10\% in excess wrt. the expected value. The variation of the PSF over the field of view, the vignetting and the plate scale have been also measured in flight using Cyg X-1 observed in nine different positions in the FOV. There results are reported in the dedicated paper by Feldman et al., this issue.

\subsubsection{Straylight}
\label{sec:straylight}
\begin{figure}[ht]
   \centering
\includegraphics[height=5cm]{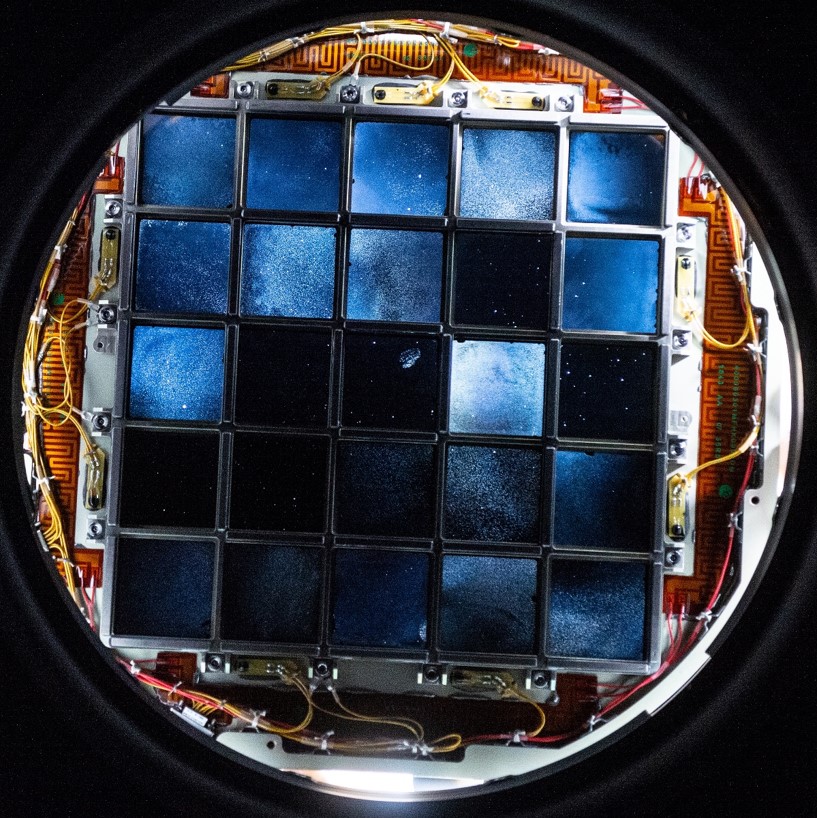}
\includegraphics[height=5cm]{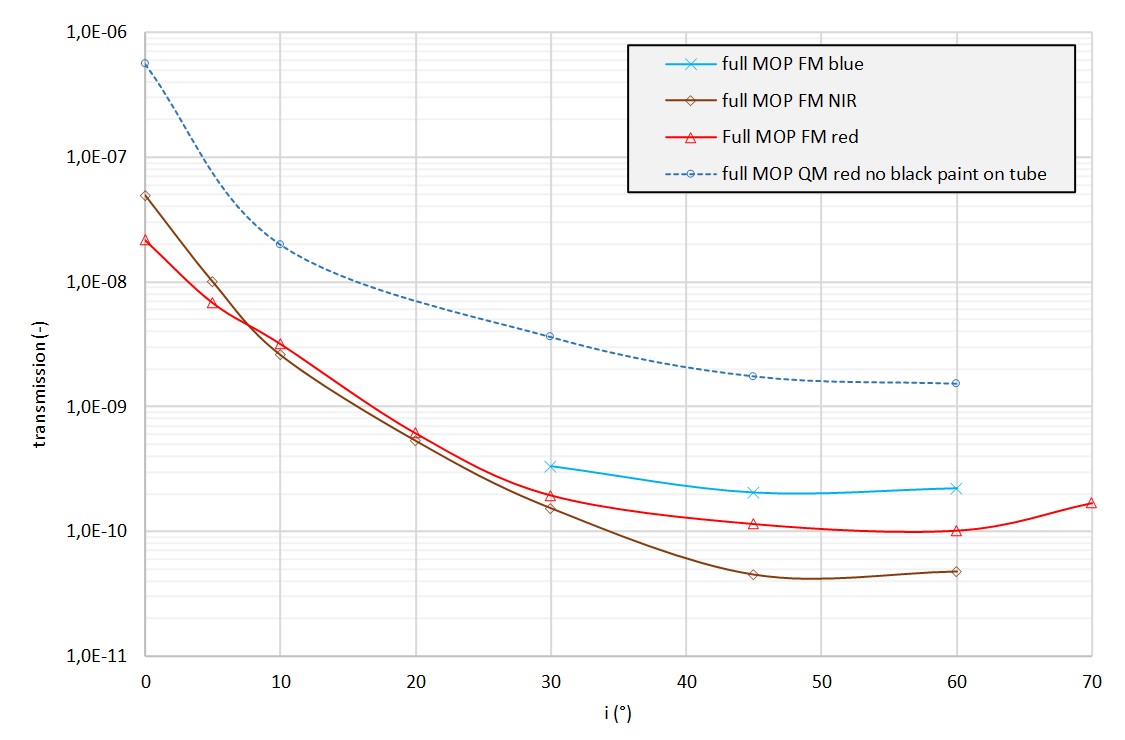}   
\caption{(Top) Back of the FM MOP illuminated by a white lamp. (Bottom) Optical transmission of the MOP.}
   \label{fig:MOP transmission}
   \end{figure}

The MXT pnCCD is sensitive to light of wavelength up to 1.1 µm, hence visible photons can pile up in a pixel and emulate a soft X-ray photon. In orbit the main source of this optical loading is the Earth. The requirement was to have a visible straylight induced background noise lower than 2$\times 10^6$ ph/cm$^2$/s with a guard angle wrt the Earth limb of 20°. Due to accomodation limitations on the platform, it was not feasible to implement a long baffle in front of MXT, and the protection relies on two aluminum (Al) layers: one in front of the MOP  (70 nm) and one in front of the CCD (100 nm). Fig. \ref{fig:MOP transmission} (top) illustrates how a residual visible light passes through the FM MOP illuminated by a white lamp. The blue color corresponds to a higher reflectivity of the Iridium layer and a  more efficient diffusion process at shorter wavelengths. The diffusion performance disparity between the MPOs is significant and the detected number of pinholes is very low  (less than 10$^{-5}$ surface fraction). 

Fig. \ref{fig:MOP transmission} (bottom) illustrates the measured averaged transmission from the MOP to a 1 cm$^2$ CCD located at the intended position of the MXT pnCCD, for three wavelengths (460, 638 and 830 nm).  $i$ is the angle wrt. to MOP optical axis. At low $i$, the transmission is dominated by the specular rays propagating in the instrument cavity and attenuated by the black paint coating (total integrated scatter, TIS=3\%) and the stiffners. At high $i$ the transmission is dominated by the diffusion at the pores output. This is by far the dominating effect for the Earth case. There is no high dependence on wavelength and on average the transmission is 1.6 10$^{-10}$ for $i$ $\geq$ 30°.

It was not possible to measure the Al transmission for the pnCCD during the MCAM development. We had to rely on an indirect estimation  from the end-to-end measurements at PANTER, launching a collimated laser (blue, red or NIR) on the central MPO. The result is an estimation of (Q.E. x $\tau_{AlCCD}$) between 3 $\times$ 10$^{-4}$ (blue) and 3 $\times$ 10$^{-2}$ (NIR), with a uncertainty close to a factor of two. This is far more than was expected from a 100 nm Al layer (10$^{-6}$). To recover margins, we tried to add a third Al filter on the filter wheel but it failed to survive the shock tests. We also considered a local baffling of each MPO, but it was too late do develop and test it safely. The only option was to rely on energy filtering on board.  Using the best available data at delivery, the simulations estimated that the requirement would be met with 55° or greater guard angle, corresponding to a filtering efficiency at the 10$^{-12/-13}$ level at high $i$. A mission analysis estimated that this would reduce the GRB observation time by 20\%. 

At MXT first light in orbit, we figured out quickly that the Earth impact was larger than expected, with the instrument put in FDIR (Fault Detection, Isolation, and Recovery) mode due to memory overflow. The on board energy filtering was not sufficient to limit the problem. Using the Moon as a calibrator and a dedicated program exploring various guard angles, we confirmed that the limit to reach the requirement is 80° corresponding to a total filtering efficiency at few 10$^{-11}$ level. After some investigations, the only possible explanation for the discrepancy with ground measurements is a degradation of the Al layers. Possibly, a small hit during the launch phase could have degraded the Al film of the MOP or the glue at the edges. The operations have been adapted to optimize the observation time by implementing a dynamically adaptive Earth guard angle, that allows MXT to observe close to the Earth limb (1°) on the night side of the orbit, and to be as close as possible to acceptable background conditions on the day side of the orbit (see Robinet et al., this issue). Despite not being able to recover the full planned availability of MXT over the orbit (45\%), this new strategy allows for an availability of about 30\%. The main scientific impact is that some GRBs cannot be observed straight after a slew, which in most of the cases hinders a precise localization by MXT, since the source is then observed on the next orbit, once the source is significantly dimmed. 
 
\subsection{The Spectral Performance}

During the first months of operation, MXT used the calibration parameters derived from its pre-launch campaign at the PANTER X-ray facility in Neuried, Germany \citep{schneider23}. However, the first in-orbit measurements of the onboard \(^{55}\mathrm{Fe}\) calibration source revealed that the Mn~K\(\alpha\) peak was shifted to lower energies by about 100~eV relative to its nominal energy, placing the instrument outside its spectral requirements. This discrepancy likely reflects combined effects of launch stresses and the intrinsic behaviour of the detector under flight conditions, necessitating a full recalibration of the instrument.

While the onboard \(^{55}\mathrm{Fe}\) source provides strong high-energy reference lines (\(> 5.9\)~keV), calibrating MXT at low energies required additional astrophysical data. For this purpose, MXT observed the Supernova Remnant Cassiopeia~A (Cas~A) between 2024-07-25 and 2024-09-28. Cas~A exhibits several bright emission lines in the 0.6--3~keV band. Because of the MXT PSF, we needed nine individual pointings, spaced by 15 arcmin, to illuminate the full detector plane, following the strategy used in the PANTER calibration campaign \citet{gotz23}.

Using the combined dataset from the onboard \(^{55}\mathrm{Fe}\) source and the Cas~A calibration observations, we derived updated gain, offset and charge transfer efficiency (CTE) parameters. The spectral performance of MXT was then evaluated using both the onboard \(^{55}\mathrm{Fe}\) source and background fluorescence lines covering 1.5--9.7~keV. Cas~A emission lines were also used to extend centroid measurements down to 0.6~keV, although their blended nature requires more sophisticated spectral modelling to extract a reliable FWHM measurement.

The results, summarised in Table~\ref{tab:mxt_perf}, show that MXT meets all spectral performance requirements: the spectral lines centroid positions remain within \(\pm 20\)~eV across the full band, and the Al~K\(\alpha\) line at 1.49~keV yields an energy resolution of \(76.2 \pm 10.8\)~eV, compliant with the 80~eV requirement at 1.5~keV at beginning-of-life. This confirms the excellent energy response of the instrument in orbit. For further details and a full description of the analysis methods, see Miguel et al.\ (this issue).

\begin{table*}[h]
\centering
\caption[]{Measured centroid energy and energy resolution for single and all events.}
\label{tab:mxt_perf}
\begin{tabular}{lcl|cc|cc}
\hline
\textbf{Line} & \textbf{Energy (eV)} & \textbf{Origin} &
\multicolumn{2}{c|}{\textbf{Centroid Energy (eV)}} &
\multicolumn{2}{c}{\textbf{FWHM (eV)}} \\
 & & & Singles & All & Singles & All \\
\hline
O VIII       & 665   & Cas A        & 649.6 $\pm$ 3.7   & 664.1 $\pm$ 3.3 & ---  & --- \\
Fe XXII      & 1053  & Cas A        & 1032.8 $\pm$ 3.2  & 1045.1 $\pm$ 2.1 & --- & ---  \\
Mg XI        & 1350  & Cas A        & 1334.7 $\pm$ 3.1  & 1346.9 $\pm$ 1.4 & --- & ---  \\
Al K$\alpha$ & 1486  & Bkg          & 1481.9 $\pm$ 5.5  & 1491.9 $\pm$ 3.8  & 62.7 $\pm$ 14.8  & 76.2 $\pm$ 10.8 \\
Si XIII      & 1852  & Cas A        & 1837.6 $\pm$ 1.0  & 1849.2 $\pm$ 5.4 & --- & --- \\
S XV         & 2449  & Cas A        & 2424.0 $\pm$ 1.8  & 2436.4 $\pm$ 1.2 & --- & --- \\
Si esc       & 4160  & $^{55}$Fe    & 4142.8 $\pm$ 2.5  & 4155.2 $\pm$ 3.0  & 132.8 $\pm$ 6.8  & 140.7 $\pm$ 3.9 \\
Mn K$\alpha$ & 5895  & $^{55}$Fe    & 5885.0 $\pm$ 0.3  & 5905.8 $\pm$ 0.7  & 145.8 $\pm$ 0.7  & 156.9 $\pm$ 0.5 \\
Mn K$\beta$  & 6490  & $^{55}$Fe    & 6477.5 $\pm$ 0.7  & 6498.5 $\pm$ 0.2  & 155.6 $\pm$ 2.0  & 167.0 $\pm$ 1.3 \\
Cu K$\alpha$ & 8041  & Bkg          & 8040.9 $\pm$ 1.5  & 8051.9 $\pm$ 0.9  & 173.6 $\pm$ 4.4  & 176.9 $\pm$ 2.8 \\
Au L$\alpha$ & 9713  & Bkg          & 9716.9 $\pm$ 2.2  & 9727.1 $\pm$ 1.7  & 214.8 $\pm$ 6.3  & 215.2 $\pm$ 4.9 \\
\hline
\end{tabular}
\end{table*}

\subsection{The MXT Effective Area}

\begin{figure}[ht!]
   \centering
   \includegraphics[width=0.5\textwidth,angle=180]{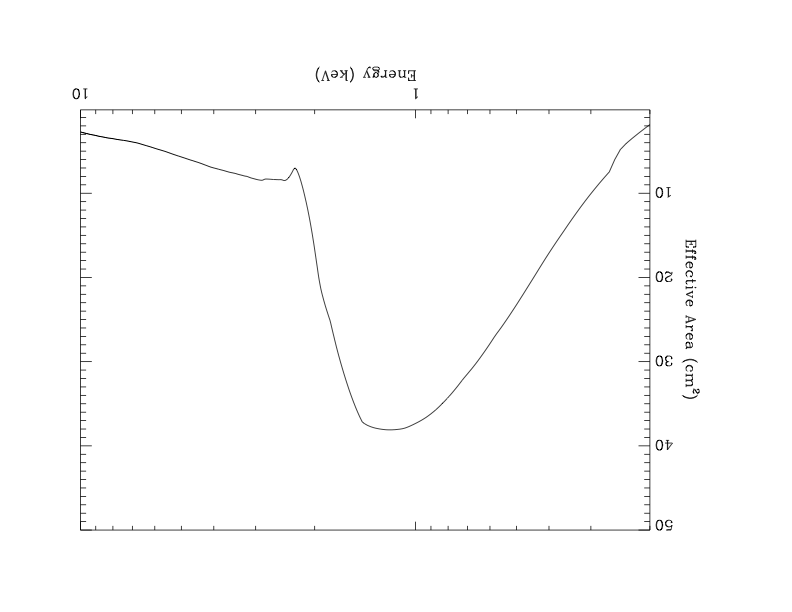}
   \caption{MXT effective area for all multiplicities.}
   \label{fig:effarea}
   \end{figure}

The MXT effective area has been modeled on ground, based on the Panter results, see Fig. 6 in \citet{gotz23}. The resulting effective area (issued from all the pixel interaction multiplicities, i.e. simples, doubles, triples and quadruples) is represented in Fig. \ref{fig:effarea}.

\begin{figure}[ht!]
   \centering
   \includegraphics[width=0.5\textwidth]{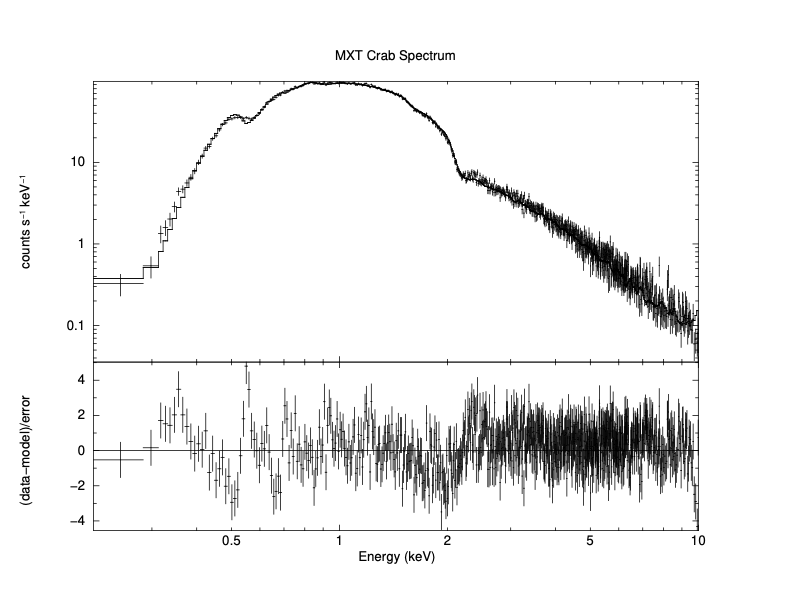}
   \caption{MXT Crab spectrum and residuals. The fit has been performed using the photoionization cross sections \citet{vern} and the abundances by \citet{wilms}. The count rate measured in the 0.2-10 keV energy range is 135 counts/s.}
   \label{fig:crab}
   \end{figure}

We observed the Crab nebula twice at the beginning of the mission on September 20$^{th}$ and 21${st}$ 2024. Here we present the first observation obtained between 21:18 U.T. and 23:20 U.T, for an effective exposure time of 2940 s. By analyzing its spectrum using the effective area obtained on ground, we compared the expected and the measured flux of the Crab nebula in MXT. The expected model flux in the 0.5-10 keV energy range is 2.7$\times$10$^{-8}$ erg cm$^{-2}$ s$^{-1}$ \cite{weisskopf10} and with MXT we measure 3.2$\pm$0.1$\times$10$^{-8}$ erg cm$^{-2}$ s$^{-1}$, in rather good agreement (18\% difference). Concerning the derived spectral parameters we obtain a photon index $\Gamma$ = 2.01$\pm$0.1, which is slightly harder than the one obtained by \citet{weisskopf10} (2.04-2.26) and an absorption column $N_H$ = 0.424$\pm0.003$ $\times$ 10$^{22}$ cm$^{-2}$, in agreement with the values reported in the same paper (0.33-0.49). The residual differences in flux and photon index, as well as some structures on the fit residuals, will need some optimization of the source extraction region as well as minor effective area updates. The Crab nebula spectrum and its fit are shown in Figure \ref{fig:crab}. The background used for this analysis is the one derived from deep observations of the Lockmann Hole region, see later.

\subsubsection{Intercalibration with XRT and FXT}

\begin{figure}
   \centering
   \includegraphics[width=0.5\textwidth]{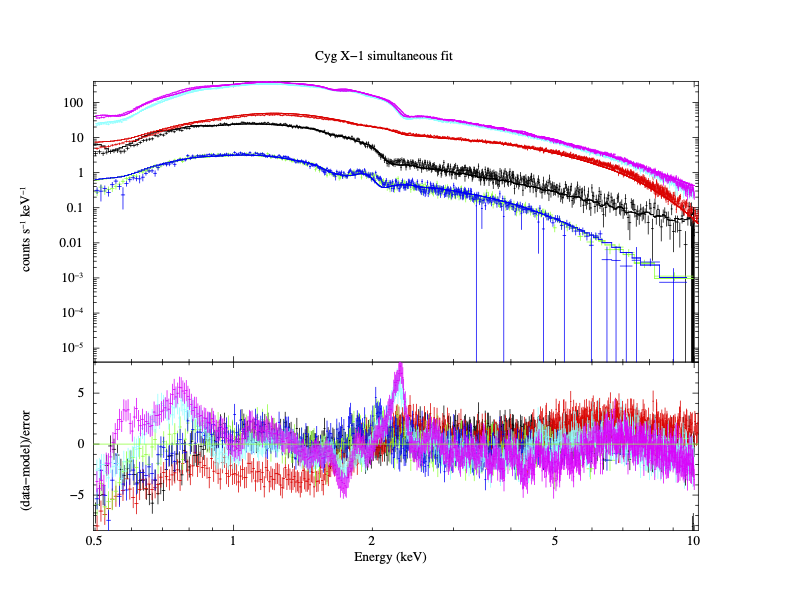}
   \caption{Simultaneous spectra of the Black Hole binary Cyg X-1 obtained during the SVOM Commissioning phase on October 20$^{th}$ 2024. In the upper panel we show the measured spectra obtained with (from top to bottom) FXT A, FXT B, XRT, MXT and WXTs. In the bottom panel the residuals with respect to the best fit model are reported. The (joint) best fit model is represented by the (absorbed) sum of a multicolor black body and a power law. The inner temperature of the disk is measured to be 0.25$\pm$0.01 (90\% c.l.) keV and the power law index 1.79$\pm$0.01 (90\% c.l.). The residuals around 2 keV are due to effective area modelling features in the FXTs.}
   \label{fig:crosscalib}
   \end{figure}

In order to further validate our effective area modelling we also performed (quasi-) simultaneous observations of Cyg X-1 using MXT, Einstein Probe WXT, Einstein Probe FXT A/B, and Swift XRT. All observations were acquired on October 20$^{th}$ 2024, and the exposure times were about 9 ks for XRT, WXT, and FXTs and about 14 ks for MXT. Thanks to this observation campaign (see Fig. \ref{fig:crosscalib}) we could accurately estimate the relative calibration constants for most of the currently flying X-ray  telescopes, and they all agree within +/- $\sim$10\% (see Table \ref{tab:intercal}), meaning that our on-ground modelling of the MXT effective area is substantially correct.

 \begin{table*}[ht]
    \centering
    \caption{Intercalibration constants between XRT, MXT, FXTs, and WXTs.}
        \begin{tabular}{ccccccc}
            \hline \hline
Telescope & MXT & XRT & FXTA & FXTB & WXT (CMOS 14) & WXT (CMOS 37) \\
\hline
Constant & 0.94 & 1.0 & 1.09 & 1.12 & 1.08 & 1.08\\
            \hline
        \end{tabular}
    \label{tab:intercal}
    \end{table*}

\subsection{The in-flight background measurements and sensitivity}

The MXT background can be decomposed into a sky and in an internal component, both of which shall be measured to produce correct spectral analyses. The sky background comprises cosmic X-ray background (CXB) and soft X-ray background originating from local structures surrounding Solar System or from diffuse sources in the Milky Way. The internal background is mainly caused by high-energy protons and neutrons interacting with the MXT camera structure.
This section characterizes the flux and spectrum of these MXT camera background components.

\subsubsection{The Internal Background}
The high-energy neutrons and protons that cause the internal background mainly originate from cosmic rays and solar wind particles trapped by the Earth's magnetic field. These particles penetrate the shielding surrounding the detector and generate secondary particles, such as X-ray photons and electrons.
To investigate the characteristics of the internal background, we collected $10^5$ seconds of in-orbit data with the MXT camera's filter wheel in the closed position during the period from August 16$^{th}$ to September 13$^{th}$, 2024. Figure \ref{fig:bkg} shows the spectrum of the internal background. The total count rate in the 0.3--10.0 keV energy range is 0.124$\pm$0.001 cts/sec. Fluorescent X-ray lines from aluminum, nickel, copper, and gold originating from the detector shielding and electronic components were identified. For more details see Moita et al. this issue.

\subsubsection{The Sky Background}
The CXB is an isotropic X-ray emission, about 85–90$\%$ of which in the 2–10 keV band has been resolved into point sources, primarily active galactic nuclei, by observations from ROSAT, Chandra and XMM-Newton \citep{2003ApJ...588..696M, moretti03}. In several keV range, the spectrum is essentially described by a power law or a broken power law. In addition to CXB, MXT may observe soft X-ray backgrounds from the Milky Way and the vicinity of the Earth. The Local Hot Bubble investigated in \cite{1995A&A...294L..25E, 2008ApJ...674..209K, 2017ApJ...834...33L} is a plasma structure non-uniformly enveloping the solar system (size: 50–200 pc; temperature: kT = 0.09–0.11 keV). It is considered to be a cavity resulting from single or multiple supernova explosions, filled with high-temperature plasma. The warm-hot circumgalactic medium refers to the relatively high-temperature portion (kT = $<$0.1 keV by \cite{2024A&A...681A..78L}) of the circumgalactic medium that fills the Milky Way's halo (size: $\sim$300 kpc), while the galactic corona is a structure with an even higher temperature (kT = 0.3–1.5 keV). 
To investigate the characteristics of the simplest sky background observed with MXT, we performed observations of the Lockman Hole (l, b = 149.77 deg, 52.03 deg), a sky region with few X-ray sources. The observations were conducted in seven periods between November 9th and December 8th, 2024, and the data from November 9th to 11th,  were excluded to avoid high Solar particle background (Kp index $>$ 3). The remaining sky observations data from November 20th to December 8th were analyzed, and the average count rate including the internal background was $(8.589 \pm 0.054)\times 10^{-1} \rm ~counts/s$ across the entire camera field with energy range 0.3-10.0 keV, for an exposure time of 29 ks. After subtracting the internal background event rate measured with the MXT camera with the filter wheel in closed position, the count rate from the Lockman hole was $(7.25 \pm 0.05)\times 10^{-1} \rm ~counts/s$ in same camera region and energy range. 

\begin{figure*}
    \centering
    \includegraphics[width=0.5\linewidth]{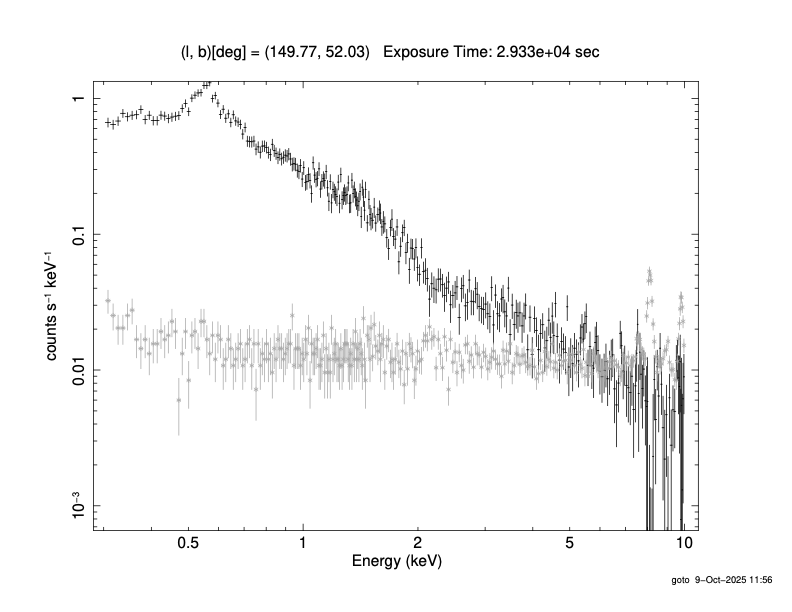}\includegraphics[width=0.45\linewidth]{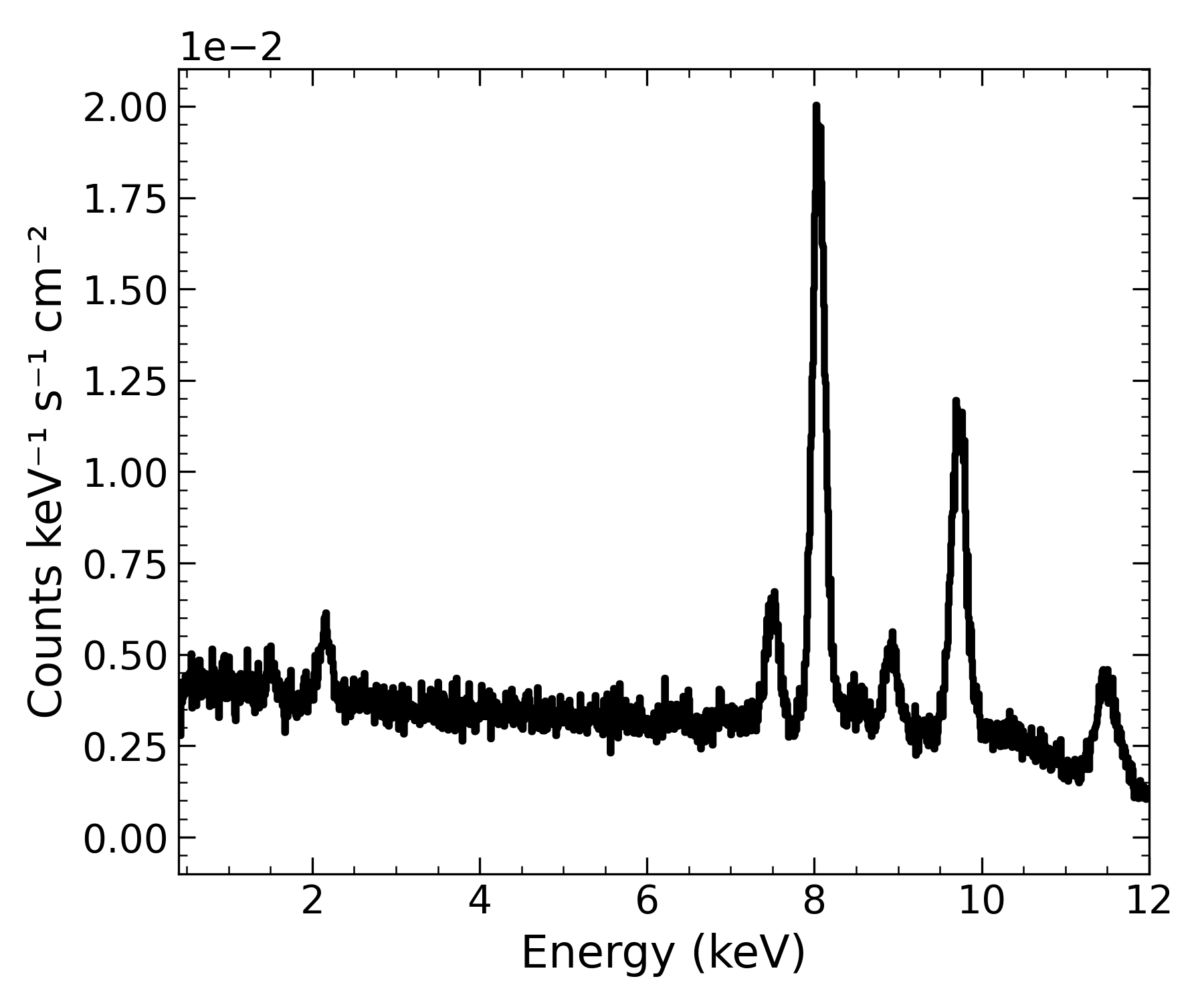}
    \caption{(Left)Black: Spectrum taken from the entire MXT camera field of view observed in the direction of the Lockman Hole at 0.3–10 keV, with internal background already subtracted. Gray plot with star marks: Spectrum of the internal background obtained at the same time as the Lockman Hole measurement. (Right): Internal MXT Background spectrum.}
    \label{fig:bkg}
\end{figure*}

Fig. \ref{fig:bkg} shows that, in addition to the continuum from CXB, a peaked structure centred around ~$0.57~\rm keV$ is observed, which is mostly likely to be the O VII $\rm K \alpha$ triplet (investigated in \cite{2001A&A...376.1113P}). This excess may persist even during periods of low geomagnetic activity; as demonstrated in \cite{2007PASJ...59S.133F,2010PASJ...62..981E}, solar wind charge exchange (SWCX) can occur even at low Kp values. OVII emission from local hot bubbles and/or the hot intergalactic medium (galactic corona) of the Milky Way is also expected (\cite{2007A&A...475..901K}). Detailed spectral analysis, temporal variation analysis, and directional dependence studies are scheduled to be carried out in the near future.



\subsubsection{MXT Sensitivity}

Taking the background and effective area values presented above into account and assuming a GRB afterglow-like spectrum we obtain a 3 $\sigma$ detection limit of about 2-3$\times$ 10$^{-12}$ erg cm$^{-2}$ s$^{-1}$ in the 0.3--10 keV energy range for a 10 ks observation, compatible with the pre-launch estimations. We note however, that the Lockman Hole observations have been selected over periods of low cosmic ray induced background. We found that during periods of high solar activity, despite the presence of an electron diverter, high energy particles can leak on to the focal plane through the optics (see Feldman et al., this issue) and locally enhance the measured background up to a factor two, thereby reducing the MXT sensitivity. Another source of background is Earth induced straylight (see \ref{sec:straylight}), but this affects mainly the energy range below 0.5 keV.

\subsection{The MXT localization capabilities}

\begin{figure}
    \centering \includegraphics[width=0.45\linewidth]{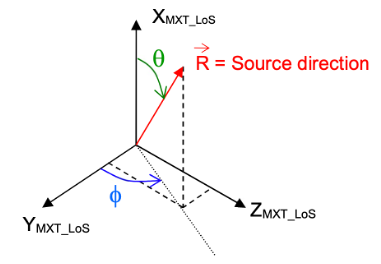}
    \caption{MXT line of sight reference definition.}
    \label{fig:rf}
\end{figure}

The SVOM main pointing reference is the VT optical axis. The bias between the VT reference frame and the satellite AOCS has been characterized at the beginning of the mission and has been consistently below 10 arc seconds during the Commissioning Phase (see Qiu et al. this issue).
In order to accurately localize the MXT sources in J2000
it has been chosen to build a MXT-VT bias rotation matrix allowing transformations between the MXT and VT reference systems. This matrix allows also to select on board the window in the VT data, 
in order to look more efficiently for optical afterglows in the VHF data, by reducing the number of potential candidate sources. In Fig. \ref{fig:rf} we show the definition
of the MXT line of sight. 

By following this formalism, and assuming MXT and VT are observing the same source one can define the bias matrix ($M_{VT,MXT}$) between the two instruments as follows (in usual spherical coordinates)

\begin{equation}
    V_{MXT}(\theta_{MXT}, \phi_{MXT})= [M_{VT,MXT}] V_{VT}(\theta_{VT}, \phi_{VT}),
\label{eq:bias}
\end{equation}

where

\begin{equation}
   V_{MXT}(\theta_{MXT}, \phi_{MXT})= \begin{bmatrix}
       cos \theta_{MXT}\\
       cos \phi_{MXT} \quad  sin \theta_{MXT}\\
       sin \phi_{MXT}  \quad sin \theta_{MXT}\\
   \end{bmatrix}
\end{equation}

and

\begin{equation}
   V_{VT}(\theta_{VT}, \phi_{VT})= \begin{bmatrix}
       cos \theta_{VT}\\
       cos \phi_{VT} \quad  sin \theta_{VT}\\
       sin \phi_{VT}  \quad sin \theta_{VT}\\
   \end{bmatrix}
\end{equation}

In order to solve Eq. \ref{eq:bias} at least three observations of the same source in different positions are needed.

During the first months of the mission, we calibrated the bias between MXT and VT line of sight (LoS) through a series of observations of the bright X-ray source Cyg X-1 in nine different positions in the FOV (including the four corners and the centre), which resulted in the computation of the first MXT/VT bias rotation matrix.
Once the bias rotation matrix is applied, the offset measured in the J2000 frame between the MXT derived position and the Cyg X-1 catalogue position is significantly reduced as can be seen in Table \ref{tab:bias}. One can see that the residual bias is always smaller than 30 arcsec, and on average 17.3 arcsec. For comparison purposes, the statistical only error for such observations would be of the order of a few arcseconds, hence systematics are the dominant source of error for such high SNR sources. This is, however, not the case for GRB afterglows that rarely reach these high SNRs, see Robinet et al., this issue.
The fact that the residual bias does not vanish completely is probably due to the fact that the transformation between the MXT and VT LoS is not a simple linear rotation, but may include some additional distortions. 

We noticed that the MXT position degraded progressively with time. So in April 2025 we performed a series of observations on GX 5-1 to produce a new bias matrix. However the accuracy did not improve. So new Cyg X-1 observations were obtained in June 2025
in five different positions (the center and the four corners).
Using these update bias matrix the new residual bias is 13 arc sec, which is an acceptable systematic error for MXT.
We plan to update the bias matrix once every three to six months.

\begin{table}[ht!]
\begin{center}
\caption[]{ Difference of the measured J2000 MXT position of Cyg X-1 with respect to its catalogue position, before and after correcting for the MXT/VT LoS bias matrix.}\label{tab:bias}


 \begin{tabular}{cccc}
 \hline\noalign{\smallskip}
Position &  SVOM      & Uncorrected Offset  & Residual bias                     \\
& OBSID &  (arc sec) & (arc sec)\\
\hline\noalign{\smallskip}
P1	&140852022	&105.3&	28.0\\
P8	&140852023	&84.1	&15.0\\
P3	&140852024	&79.7	&15.8\\
P0	&140852025	&77.2	&7.9\\
P4	&140852077	&92.4	&12.1\\
P5	&140852078	&81.1	&14.8\\
P0 	&140852118	&87.9	&12.8\\
P7	&140852119	&99.4	&19.3\\
P2	&140852120	&105.1	&30.2\\
P6	&140852121	&101.5	&17.4\\
\noalign{\smallskip}\hline
\end{tabular}
\end{center}
\end{table}

The knowledge of the bias allowed MXT to provide accurate GRB afterglow positions for tens of objects with absolute accuracies smaller than 2 arc minutes. For more details, see Robinet et al. this issue.

\section{Conclusions}

We presented the scientific performance of the MXT instrument as measured
during the SVOM Commissioning phase. We have shown that the pre-launch requirements are met in terms of sensitivity, spectral resolution, optical and localization capabilities. 
However the instrumental availability is below the expected one with a  $\sim$25-30\% loss, due to Earth straylight that is not efficiently filtered over part of the SVOM orbit. The implementation of a baffle in front of the MXT optics would have mitigated the problem, but the volume limitations on the satellite have not allowed it.

MXT is hence able to quickly localize and characterize GRB afterglows in terms of timing and spectral properties, especially over the first orbit after trigger.

\begin{acknowledgements}
The Space-based multi-band astronomical Variable Objects Monitor (SVOM) is a joint Chinese-French mission led by the Chinese National Space Administration (CNSA), the French Space Agency (CNES), and the Chinese Academy of Sciences (CAS). We gratefully acknowledge the unwavering support of NSSC, IAMCAS, XIOPM, NAOC, IHEP, CNES, CEA, and CNRS.
\end{acknowledgements}



\bibliography{bibtex}

\label{lastpage}

\end{document}